\documentclass[prb,twocolumn,amsmath,amssymb,aps,showpacs,floatfix]{revtex4-1}

\usepackage{graphicx}
\usepackage{dcolumn}
\usepackage{bm}
\usepackage{caption}
\usepackage{subcaption}
\usepackage{xcolor}
\usepackage{caption,subcaption}
\begin{document}

\preprint{APS/123-QED}

\title{Non-universal Critical Aging Scaling in Three-dimensional Heisenberg Antiferromagnets}

\author{Riya Nandi} \email{riya11@vt.edu}
\author{Uwe C. T\"auber} \email{tauber@vt.edu}
\affiliation{Department of Physics (MC 0435) and Center for Soft Matter and Biological Physics, \\ 
             Virginia Tech, Robeson Hall, 850 West Campus Drive, Blacksburg, Virginia 24061, USA}
\date{\today}

\begin{abstract}
We numerically investigate the stationary and non-equilibrium critical dynamics in three-dimensional isotropic Heisenberg antiferromagnets. 
Since the non-conserved staggered magnetization couples dynamically to the conserved magnetization density, we employ a hybrid simulation algorithm that combines reversible spin precession with relaxational Kawasaki spin exchange processes. 
We measure the dynamic critical exponent and identify a suitable intermediate time window to obtain the aging scaling exponents.  
Our results support an earlier renormalization group prediction:
While the critical aging collapse exponent assumes a universal value, the associated temporal decay exponent in the two-time spin autocorrelation function depends on the initial distribution of the conserved fields; here, specifically on the width of the initial spin orientation distribution.
\end{abstract}




\maketitle

\section{Introduction} 
Systems near a continuous phase transition or critical point display many intriguing collective properties such as large response to even minute perturbations, long-range correlations that extend far beyond any microscopic interaction range, and the concurrent critical slowing-down of characteristic time scales that become much larger than inverse microscopic relaxation rates (for overviews, see, e.g, Refs.~\cite{Ma1967book, HohenbergHalperin1977theory, Janssen1979fieldtheory, ChaikinLubensky1995book, Cardy1996book, Vasilev2004book, FolkMoser2006review, Kamenev2011book, Tauber2014critical}).
Indeed, singularities in both thermodynamic properties as well as transport coefficients may be viewed as originating from the system's diverging correlation length, typically following a power law $\xi(\tau) \sim |\tau|^{-\nu}$ with critical exponent $\nu$ as function of the distance $\tau$ from the critical point, e.g., $\tau \propto T - T_c$ for a continuous equilibrium phase transition at critical temperature $T_c$.
Similarly, mesoscopic relaxation times tend to scale as $t_c(\tau) \sim \xi(\tau)^z \sim |\tau|^{-z \nu}$, with dynamic critical exponent $z$.
As the governing collective length and time scales diverge at the critical point, systems with infinitely many degrees of freedom acquire scale invariance as an emergent new symmetry, which is reflected in non-trivial scaling laws for physical observables that mathematically assume the form of generalized homogeneous functions.
Yet in a finite system, say with sufficiently large linear extension $L$, the characteristic relaxation time should follow the finite-size scaling law $t_c(\tau,L) = \xi(\tau)^z \, {\hat t}(\xi/L)$, with ${\hat t}(0) =$ const., while the scaling function at large arguments ${\hat t}(y \gg 1) \sim y^{-z}$ in order to eliminate the $\tau$ dependence as $\tau \to 0$, implying $t_c(0,L) \sim L^z$.

The dominance of a critical system's order parameter correlation length and relaxation time over microscopic length and time scales is also at the core of the remarkable universality properties of continuous phase transitions, both near and far from thermal equilibrium:
Quite distinct continuous phase transitions driven by entirely different microscopic mechanisms are characterized by identical static critical exponents, and may hence be subsumed into broad universality classes that  for short-range microscopic interactions only depend on the underlying symmetry of the order parameter and the system's dimensionality $d$.
Beyond the upper critical dimension $d_c$, even the $d$ dependence disappears, and the critical exponents may be obtained by means of straightforward mean-field approximations; for $d < d_c$, critical fluctuations dominate and cause marked deviations from mean-field predictions.
However, physical systems that in thermal equilibrium fall into the same static universality class, may still display distinct dynamical critical scaling behavior: 
Crucial differences arise if the order parameter is conserved under the dynamics or not; and if it is coupled to other conserved slow modes \cite{HohenbergHalperin1977theory, ChaikinLubensky1995book, Cardy1996book, Vasilev2004book, FolkMoser2006review, Tauber2014critical}.

Nevertheless, despite the inevitable split of static into varied dynamic universality classes, quite distinct physical systems may be described by the same dynamical critical exponent; e.g., this is the case for planar ferromagnets (model E in Hohenberg and Halperin's classification scheme \cite{HohenbergHalperin1977theory}), superfluid Helium 4 (model F), and isotropic antiferromagnets (model G), for which $z = d / 2$ in $d \leq d_c = 4$ spatial dimensions \cite{HalperinHohenbergSiggia1976, GuntonKawasaki1976, FreedmanMazenko1976}.
Indeed, these systems may be subsumed into one $O(n)$-symmetric dynamical universality class in which an $n$-component non-conserved order parameter ($n = 2$ for models E and F, $n = 3$ for model G) is reversibly coupled to $n (n-1) / 2$ non-critical conserved fields (i.e., a scalar for model E, a three-component vector field for model G) \cite{SasvariSchwablSzepfalusy1975, SasvariSzepfalusy1977, DominicisPeliti1978}.

A thorough theoretical understanding and a systematic classification of dynamical systems near criticality have been gained through extensive numerical simulations \cite{LandauKrech1999spin,ChenLandau1994spin}, and by means of the dynamic renormalization group approach which crucially exploits the emerging scale invariance \cite{Ma1967book, HohenbergHalperin1977theory}, especially its field-theoretic variant \cite{Janssen1979fieldtheory, Cardy1996book, Vasilev2004book, FolkMoser2006review, Kamenev2011book, Tauber2014critical}.
This extends beyond the steady-state kinetics to the critical non-equilibrium relaxation regime; here, the system is initially prepared in a disordered configuration with prescribed (usually Gaussian) probability distribution, and is subsequently quenched to the critical point, i.e.,  forced to evolve under critical parameter values.
Hence it can only relax to stationarity algebraically slowly and retains memory to the initial state, resulting in the breaking of time translation invariance and associated characteristic aging scaling behavior for the spatial two-time order parameter correlation function 
\begin{equation}
  C(t,s,r,\tau) = r^{- (d - 2 + \eta)} \, (t / s)^{\theta - 1} \ 
  {\hat C}(r / \xi, t / \xi^z)
\label{cagscf}
\end{equation}
in the limit $t \gg s$, with the standard static Fisher exponent $\eta$ and the initial-slip exponent $\theta$ \cite{JanssenSchaubSchmittmann1989, Janssen1992review, CalabreseGambassi2005, HenkelPleimling2010book, Tauber2014critical}.
For the autocorrelations ($r = 0$) at criticality ($\tau = 0$), this turns into a simple aging scaling form
\begin{equation}
  C(t,s) \sim s^{-b} \, (t / s)^{-\lambda / z}
\label{sagscf}
\end{equation}
with the aging collapse exponent $b = (d - 2 + \eta) / z$ and the autocorrelation exponent $\lambda = d - 2 + \eta + z (1 - \theta)$.
Importantly, therefore, the dynamic exponent $z$ can be directly inferred from the early-time relaxation scaling; one thus need not necessaily contend with the long relaxation times to reach the quasi-stationary regime.

For $O(n)$-symmetric systems with conserved order parameter, $\lambda = d + 2$ exactly \cite{Sire2004autocorrelation}, implying the exact scaling relation $\theta = 1 - (4 - \eta) / z$.
For purely irreversible diffusive relaxation (model B) with $z = 4 - \eta$, consequently $\theta = 0$ \cite{JanssenSchaubSchmittmann1989}; and similarly for isotropic ferromagnets (model J) whose dynamics entails a reversible spin precession term, $z = (d + 2 - \eta) / 2$ for $d \leq 6$, whence $\theta = - (6 - d - \eta) / (d + 2 - \eta) < 0$ \cite{OerdingJanssen1993non}.
In contrast, if the order parameter is not conserved, $\theta$ in general represents an independent critical exponent. 
It assumes a universal value  for purely dissipative relaxational kinetics (model A) that can for instance be computed perturbationally in a dimensional $\epsilon = d_c - d$ expansion; to first order $\theta = (n + 2) \, \epsilon / 4 (n + 8) + O(\epsilon^2)$ for $d \leq d_c = 4$ \cite{JanssenSchaubSchmittmann1989, Janssen1992review}.

Yet in their one-loop renormalization group analysis for the critical dynamics of a non-conserved order parameter that is subject to reversible mode coupling to a non-critical conserved field, based on a continuum description in terms of coupled non-linear Langevin equations \cite{SasvariSzepfalusy1977, DominicisPeliti1978, Tauber2014critical}, Oerding and Janssen found in 1993 that the initial-slip exponent $\theta$ explicitly depends on the Gaussian distribution width $c_0$ for the magnitudes of the initial values of the conserved mode; quite surprisingly therefore, they predicted non-universal critical aging scaling for this universality class \cite{OerdingJanssen1993non}.
Indeed, in their analysis, $z \, \theta = - \gamma_\lambda - \eta_0 / 2$ is determined by the anomalous scaling dimension of the order parameter relaxation rate $\gamma_\lambda = \epsilon / 2$ (which holds to all orders at the strong dynamic scaling fixed point) and of the fields on the initial-time surface $\eta_0 = \left( 3 c_0 / 2 - 16 / 11 \right) \epsilon + O(\epsilon^2)$ (for $n = 3$), which increases linearly with the width $c_0$.
To our knowledge, this striking prediction has not yet been tested either numerically or experimentally to date.

In this work, we provide numerical evidence, gathered through a hybrid simulation algorithm for the critical dynamics of isotropic antiferromagnets on a three-dimensional lattice which combines reversible spin precession with relaxational Kawasaki spin exchange processes, that lends support to this remarkable assertion:
We demonstrate that the spin autocorrelation exponent $\lambda$ depends on the width of the initial spin orientation distribution, whereas we measure a universal value b for the aging scaling exponent. 

\section{Model and simulation algorithm}
We consider the isotropic Heisenberg model (in the absence of an external field) with the Hamiltonian 
\begin{equation}
  \mathcal{H} = - J \sum_{< i j >} \vec{S}_i \cdot \vec{S}_j
\label{heisham}
\end{equation}
governing the nearest-neighbor exchange interactions of three-component vector spin variables ${\vec S}_i$ on a three-dimensional simple cubic lattice of linear size $L$ (with $L$ up to $70$ lattice sites) with with periodic boundary conditions.
For both ferro- ($J > 0$) and antiferromagnetic ($J < 0$) exchange couplings, a continuous transition from a disordered high-temperature paramagnetic state to the ordered phase occurs for $L \to \infty$ at $k_{\rm B} T_c^\infty \approx 1.446\,J$ in three dimensions, which belongs to the static universality class of the $O(3)$-symmetric Ginzburg--Landau--Wilson vector model with Fisher exponent $\eta \approx 0.0355$ \cite{ZinnJustin2001}.
For classical Heisenberg magnets, this has been confirmed extensively through high-resolution Monte Carlo simulations \cite{Peczak1991high, Chen1993static}. 
Sophisticated methods such as single-cluster updates and histogram reweighing, optimization techniques, and finite-size scaling have provided very accurate estimates of the static critical exponents \cite{Holm1993critical, Chen1993static}.  

In the absence of an external field, the Hamiltonian (\ref{heisham}) is rotationally symmetric in spin space, and hence the components of the total magnetization $M^\alpha = \sum_i S_i^\alpha$ ($\alpha = x,y,z$) are conserved under the dynamics: $\{ \mathcal{H}, M^\alpha \} = 0$; here we employ a classical notation in terms of Poisson brackets, following the standard correspondence with quantum-mechanical commutators. 
For an antiferromagnet ($J < 0$), the order parameter is however represented by the components of the non-conserved staggered magnetization vector $N^\alpha$, with alternating $\pm 1$ signs affixed to adjacent lattice sites. 
At zero temperature, the microscopic spin variables at each lattice site $i$ obey the coupled Heisenberg equations of motion $d S_i^\alpha(t)/dt = \{ \mathcal{H} , S_i^\alpha(t) \}$, where the spin vector components satisfy the standard angular momentum Poisson brackets $\{ S_i^\alpha , S_j^\beta \} = \sum_\gamma \epsilon^{\alpha\beta\gamma} \, S_i^\gamma \, \delta_{ij}$ with the fully antisymmetric unit tensor $\epsilon^{\alpha\beta\gamma} = \pm 1$ for $(\alpha,\beta,\gamma) = (x,y,z)$ and $(z,y,x)$, respectively, and cyclic permutations of these coordinate indices.
Hence one obtains $3 L^3$ coupled deterministic equations of motion representing the spin vectors precessing in their local effective fields \cite{MaMazenko1974}, 
\begin{equation}
  \frac{d \vec{S}_i(t)}{dt} = 
  \vec{S}_i(t) \times \frac{\partial \mathcal{H}}{\partial \vec{S}_i(t)} \ .
\label{heiseqm}
\end{equation} 
Numerically, these reversible equations of motion for individual spins are readily integrated simultaneously \cite{LandauKrech1999spin, ChenLandau1994spin}, e.g., using fourth-order predictor-corrector methods.

At non-zero temperature, one must supplement this dynamics with an appropriate relaxational kinetics that however preserves the total magnetization.
One efficient means is to employ Kawasaki Monte Carlo kinetics, wherein the lattice configuration is stochastically updated by exchanging neighboring spins subject to standard Metropolis rules \cite{Kawasaki1966}. 
Yet the conservation law for the total magnetization is satisfied only within truncation error bounds set by the numerical integration scheme; in addition, one needs to adequately balance the integration time steps with a suitable number of concurrent Kawasaki spin exchange processes.
We determined the integration time increment $\Delta t = 0.01 / J$ to be optimal, with each numerical integration step separated by ten Monte Carlo sweeps over the entire lattice; we shall refer to this combination as one simulation time step (STS). 

\section{Numerical results}
\subsection{Dynamical critical exponent}
The dynamical critical exponent for isotropic Heisenberg ferromagnets $z \approx 2.48$ \cite{ChenLandau1994spin, Murtazaev2003investigation} and antiferromagnets $z = 1.5$ \cite{BunkerChenLandau1996critical, TsaiLandau2003} in three dimensions were numerically confirmed by means of such hybrid algorithms with remarkable accuracy, despite the major challenges imposed by the inevitable critical slowing-down, i.e., rapidly increasing relaxation time upon approaching criticality. 
In these studies, the temporal evolutions of the spins were determined by integrating the coupled equations of motion, but observations were made only on equilibrium configurations.
Alternatively, one may probe dynamical critical behavior effectively in the earlier non-equilibrium relaxation regime following an instantaneous temperature quench from $T \gg T_c$ to the critical point \cite{JanssenSchaubSchmittmann1989, Janssen1992review, CalabreseGambassi2005, HenkelPleimling2010book, Tauber2014critical}.

\begin{figure}
\includegraphics[width = 0.9 \columnwidth]{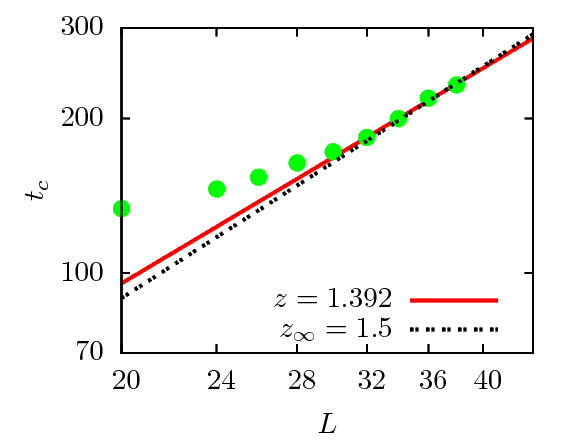}
\caption{Double-logarithmic plot of the relaxation time $t_c(L)$ vs. linear system sizes $L = 20 \ldots 38$. 
         The data points approach the exact asymptotic value $z = 1.5$ only for large $L$. 
         The best fit for the last four data points ($L \geq 32$) yields $z = 1.392(10)$. 
		 (Statistical error bars are smaller than the symbol sizes.)}
\end{figure}
In finite systems near $T_c^\infty$, the behavior of the stationary autocorrelation function can be approximated by an exponential decay $C(t) \sim e^{- t / t_c(L)}$ \cite{Abe1968}, where the relaxation time $t_c(L) \sim L^z$ diverges with system size with the dynamic critical exponent $z$. 
We extracted the relaxation time from our $C(t)$ data for different linear system sizes $L$ (Fig.~1), and found $z = 1.392(10)$ for $L \geq 32$ which is close to the exact asymptotic value $z = 1.5$ for model G.       
  
\begin{figure*}[ht]
\includegraphics[width = \columnwidth]{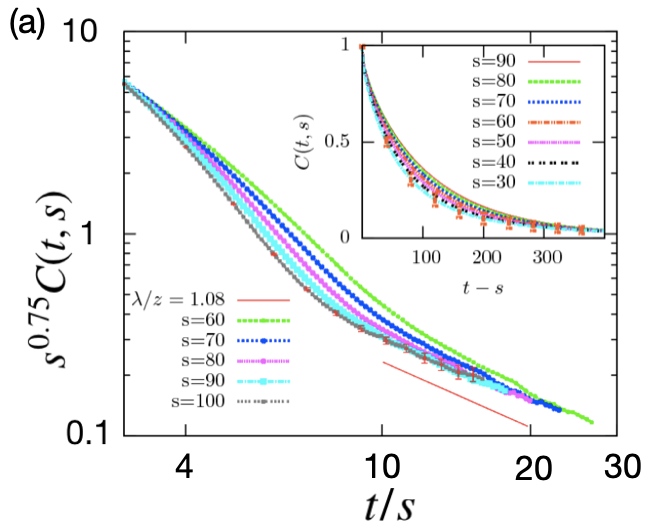} 
\includegraphics[width = \columnwidth]{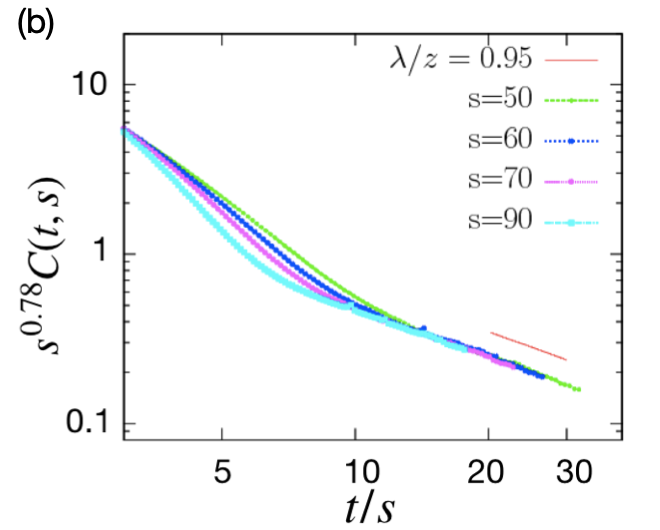}
\caption{(a) Aging scaling plots for the two-time spin autocorrelation function $C(t,s)$ following a quench to the critical temperature $T_c^\infty$ on a cubic lattice with $70^3$ sites. 
         The data for each graph were averaged over $700$ independent realizations. 
         Double-logarithmic rescaled plots for $C(t,s)$ at different waiting times $s$ collapse with the aging exponent $b = 0.75 \pm 0.04$; the corresponding decay exponent for a uniform initial distribution is $\lambda / z = 1.08 \pm 0.05$. 
         Inset: unscaled spin autocorrelations as function of $t - s$ for various $s$, demonstrating broken time translation invariance.
         Statistical errors are indicated (inset: graph for $s = 30$, main plot: graph for $s=100$) and are of similar size for all reported spin autocorrelations.
		 (b) Double-logarithmic plot of $C(t,s)$ as function of the ratio $t / s$ for various waiting times $s$ for a smaller simulation domain with $L = 40$.} 
\end{figure*}

\subsection{Critical aging scaling}
In our simulations, we used truncated Gaussian distributions of varying widths $\sigma$ for the orientations (on the three-dimensional unit sphere) of the initial spin configurations, and observed the decay of the two-time spin autocorrelation function 
\begin{equation}
  C(t,s) = L^{-3} \sum_i \left[ \langle S_i(t) \, S_i(s) \rangle - \langle S_i \rangle^2 \right] ;
\label{autocor}
\end{equation}
note that this quantity measures the autocorrelations for both the staggered magnetization order parameter and the conserved magnetization. 
We performed critical quenches from a disordered high-temperature initial configuration to $k_B T_c^\infty / J = 1.446$ on a lattice of $70^3$ spins. 
We observed an aging scaling time window for waiting times $s = 30 \ldots 100$ STS, which allowed us to extract the aging scaling exponents $b$ and $\lambda / z$ by collapsing the non-equilibrium relaxation data for $C(t,s)$ according to Eq.~(\ref{sagscf}). 
A flat orientation distribution with large $\sigma$ pertains to the setup in the renormalization group analysis of Ref.~\cite{OerdingJanssen1993non}, which instead assumed a Gaussian distribution of width $c_0$ for the magnitudes of the magnetization vector field.
We note that while our Heisenberg spin model simulations were performed at fixed ${\vec S}_i^2 = 1$, coarse-graining over sufficiently many spins with randomized orientations over a finite lattice volume would also induce a non-zero width $c_0 > 0$ for the magnitude distribution of the resulting continuous local magnetization density. 

For a uniform initial orientation distribution, we obtain a collapse exponent $b = 0.75 \pm 0.04$ for the two longest waiting times in our simulation as shown in Fig.~2(a), while the data for shorter waiting times $s$ start collapsing at this value near the large $t/s$ tail of the graph. 
It is arguable from the data that if both longer waiting and observation times were accessible to us, we would have observed a clearer scaling collapse. 
This is evident from the aging scaling data for smaller linear system size $L = 40$ displayed in Fig.~2(b), where one discerns a much better data collapse already at smaller $t/s$ ratio. 

While the asymptotic aging exponent should be $b \approx 0.69$, for our system with $L = 70$ we observe a larger value from which we infer $z = (1 + \eta) / b \approx 1.38$ in good agreement with the corresponding dynamic exponent extracted from the simulation data in Fig.~1. 
To perform a more systematic finite system size analysis, we plot the exponent $b(L)$ for four different simulation samples with linear sizes $L = 40, 60, 70$, and $100$ vs. $1 / L$ and linearly extrapolate the graph to estimate the asymptotic $b_\infty$ in the thermodynamic limit ($L \to \infty$), as displayed in Fig. ~3(a). 
Thus we obtain $b_\infty = 0.689 \pm 0.02$ which is in excellent agreement with the theoretical prediction. 
We have carried out a similar extrapolation in Fig. ~3(b) for the autocorrelation decay exponent $\lambda / z$ to estimate its asymptotic value $1.205 \pm 0.03$. 

\begin{figure*}[ht]
    \includegraphics[width= \columnwidth]{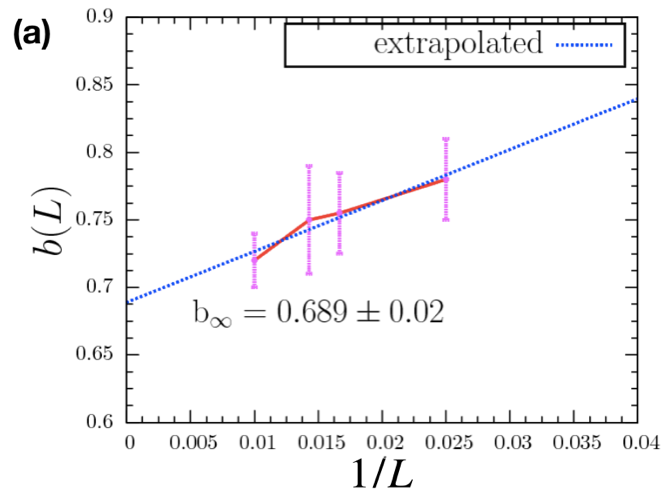}
    \includegraphics[width= \columnwidth]{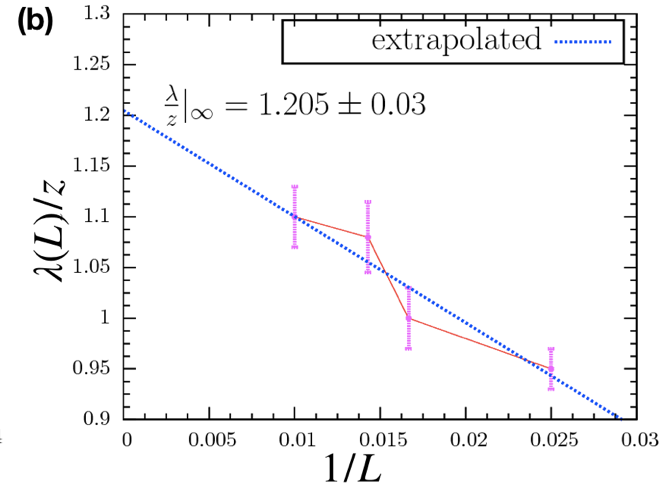}
    \caption{Finite-size (a) aging scaling exponents $b$ and (b) autocorrelation decay exponents $\lambda / z$ plotted as functions of $1 / L$ for linear system sizes $L = 40, 60, 70, 100$.
	The asymptotic values in the thermodynamic limit $L \to \infty$ resulting from linear extrapolation are $b_\infty = 0.689 \pm 0.02$ and $(\lambda / z)_\infty = 1.205 \pm 0.03$.}
\end{figure*}

\begin{figure*}[ht] 
\includegraphics[width =  \columnwidth]{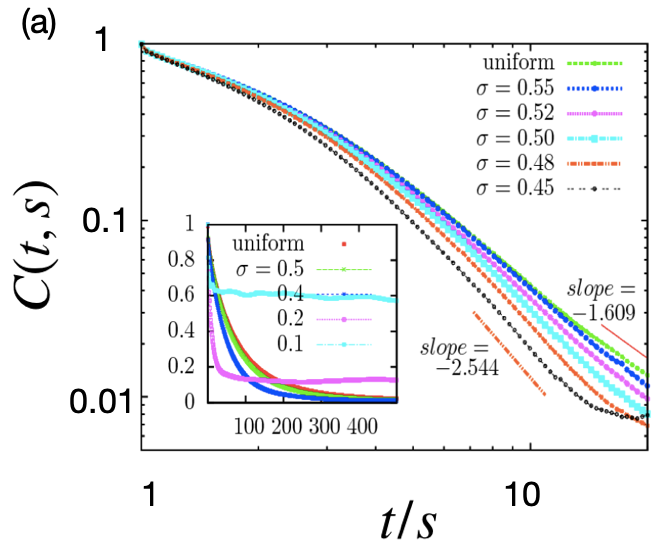} 
\includegraphics[width =  \columnwidth]{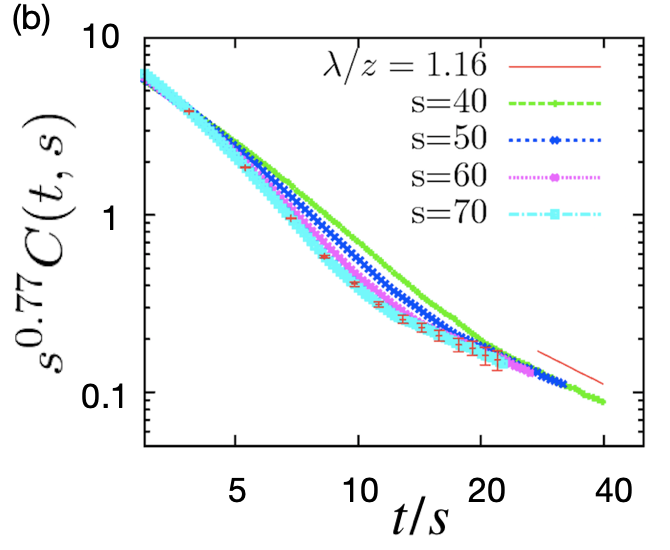}
\caption{Double-logarithmic plots of $C(t,s)$ as function of the ratio $t / s$ for  (a) different initial distribution widths $\sigma$ at fixed waiting time $s$, demonstrating that the algebraic long-time decay is non-universal.
         Inset: time evolution of the two-time spin autocorrelation function $C(t,s)$ vs. $t - s$ for different widths $\sigma$ of the initial truncated Gaussian spin orientation distribution. 
         The data for narrow $\sigma$ show distinct temporal oscillations.
        (b) for initial orientation distribution width $\sigma = 0.55$, yielding a collapse exponent $b = 0.77 \pm 0.02$ and decay exponent $\lambda(\sigma) / z = 1.16 \pm 0.07$. Statistical errors are indicated in the graph for $s=70$.
         } 
\end{figure*}
For narrow distributions ($\sigma < 0.4$), we observe transient oscillations and very slow decay (Fig.~4(a) inset). 
We attribute these features to the formation of metastable locally ordered domains; a small range of initial orientations stabilizes the domains in a finite system and prevents large-scale relaxation processes. 
But for broader distributions, we clearly detect non-universal relaxation as characterized by varying exponents $\lambda(\sigma)$ and hence $\theta(\sigma)$ (Fig.~4(a), main panel); as listed in Table~I, we observe $\lambda(\sigma)$ to decrease and $\theta(\sigma)$ to increase with growing width. 
In contrast, within our systematic and statistical errors we find identical values for the aging collapse scaling exponent $b \approx 0.75$ for all implemented orientation distribution widths; Fig.~4(b) exemplifies this for $\sigma = 0.55$. 
\begin{table}
\caption{The critical aging collapse, decay, and initial-slip exponents $b$, $\lambda / z$, and $\theta$ measured in numerical simulations for L=70 using Gaussian initial spin orientation distributions with varying width $\sigma$.}
\begin{tabular} {|c|c|c|c|}
\hline
$\sigma$ & $b$ & $\lambda / z$ & $\theta$ \\
\hline
0.5 & 0.73(3) & 1.503(70)& 0.663 \\
\hline
0.52 & 0.75(5) & 1.334(30) & 0.785 \\
\hline
0.55 & 0.77(2) & 1.16(7) &  0.910 \\
\hline 
uniform & 0.75(4) & 1.08(5) &  0.955\\
\hline 
\end{tabular}
\end{table}

As is evident from the above table, the value of exponents $\lambda(\sigma)/z$ for the narrower spin orientation distributions is much higher than the asymptotic value for this exponent for uniform distributions. 
Hence we confirm that this observation indeed clearly indicates non-universal behavior and does not result from finite-size effects. 

\section{Concluding remarks}
In summary, we have employed a hybrid numerical algorithm that combines reversible spin precession with relaxational kinetics implemented through Kawasaki spin exchange processes to study the critical dynamics of isotropic Heisenberg antiferromagnets on a three-dimensional simple cubic lattice.
Within systematic errors, our data for the characteristic spin autocorrelation relaxation time are in agreement with the dynamic critical exponent $z = 3/2$ for the model G universality class, where a non-conserved three-component order parameter (here, the staggered magnetization) is dynamically coupled to a conserved non-critical vector field (the magnetization).
Furthermore, we have investigated the non-equilibrium relaxation kinetics of isotropic antiferromagnets following a critical quench. 
We obtain a universal value $b \approx 0.75$ for the critical aging scaling collapse exponent, which within the errors of our simulation as reflected in the measurement of $z$ conforms with the asymptotic value $b \approx 0.69$. 
We have also performed a finite-size extrapolation analysis with four different linear system sizes, which yielded the correct asymptotic value of the critical aging scaling exponent $b \approx 0.689$.

Additionally, we have demonstrated that for isotropic antiferromagnets, the autocorrelation decay $\lambda / z$ and hence the initial-slip exponent $\theta$ intriguingly depend on the width $\sigma$ of the Gaussian distribution for the initial spin orientations, and hence represent non-universal critical scaling exponents. 
This remarkable and quite unusual feature was predicted by Oerding and Janssen \cite{OerdingJanssen1993non}; in a one-loop perturbative dynamical renormalization group analysis, they found an explicit dependence of the critical initial-slip exponent $\theta$ on the width of the initial Gaussian distribution width $c_0$ for the magnitude of the magnetization vector.
Our present numerical study that utilized different initial Gaussian distributions for the spin vector orientations thus validates this intriguing and unusual prediction; in accord with Ref.~\cite{OerdingJanssen1993non}, we furthermore obtain universal values for $\lambda$ and $\theta$ in the limit of sufficiently wide orientation distributions.
It would be interesting to characterize this striking non-universal behavior further by utilizing non-Gaussian initial spin distributions.
The efficient numerical simulation technique described above could also be used to explore critical short-time dynamics of more complex anisotropic antiferromagnets subject to external fields, which exhibit several distinct phase transition lines and hence allow for the presence of multi-critical points \citep{LandauBinder1978phase}.      

In closing we note that we have performed similar dynamical simulations for isotropic Heisenberg ferromagnets on a three-dimensional lattice.
However, probably owing to the much slower critical relaxation with dynamic exponent $z = (5 - \eta) / 2 \approx 2.48$, we were not able to access a sufficiently large aging time window before finite-size effects began to dominate and cut off the expected universal power laws
with the universal model J scaling exponents $b = 2 (1 + \eta) / (5 - \eta) \approx 0.42$, $\lambda = 5$, and $\theta = - (3 - \eta) / (5 - \eta) \approx - 0.60$ in three dimensions.  

\acknowledgements{
We would like to thank Harsh Chaturvedi, Sheng Chen, Shengfeng Cheng, Ruslan Mukhamadiarov, Michel Pleimling, and Priyanka for fruitful discussions and valuable suggestions that aided us in this project. 
Research was sponsored by the U.S. Army Research Office and was accomplished under Grant Number W911NF-17-1-0156. 
The views and conclusions contained in this document are those of the authors and should not be interpreted as representing the official policies, either expressed or implied, of the Army Research Office or the U.S. Government. 
The U.S. Government is authorized to reproduce and distribute reprints for Government purposes notwithstanding any copyright notation herein.
} 


\end{document}